\documentclass[onecolumn,11pt, a4,superscriptaddress,notitlepage,preprintnumbers,aps,longbibliography]{article}
\usepackage{graphicx}
\usepackage{typearea}
\usepackage{amsmath}
\usepackage{subcaption}
\usepackage[noblocks]{authblk}
\usepackage{letltxmacro,xpatch}
\usepackage[colorlinks=true,
allcolors=blue,linktocpage=true]{hyperref}
\usepackage[switch]{lineno}
\usepackage{physics}
\usepackage{color}

\usepackage{bm}
\usepackage{amssymb}
\usepackage{amsfonts}

\newcommand{\myvec}[1]{
#1
}
\newcommand{\argmin}[1]{
\mathrm{arg} \min_{#1}
}

\newcommand{\expectationl}[1]{
\langle \mathcal{O}(\myvec{x})\rangle_{#1}
}

\newcommand{\defexpectationl}[0]{
\expectationl{U}
}

\newcommand{\expectation}[1]{
\langle \mathcal{O}\rangle_{#1}
}

\newcommand{\defexpectation}[0]{
\expectation{U}
}

\newcommand{\sgn}[1]{
    \mathrm{sgn}(#1)
}
\typearea{18}

 \title{Generative quantum combinatorial optimization by means of a novel conditional generative quantum eigensolver}

\author[1]{Shunya Minami\thanks{s-minami@aist.go.jp}}
\author[2,3]{Kouhei Nakaji}
\author[1]{Yohichi Suzuki}
\author[3,4]{Al\'{a}n Aspuru-Guzik}
\author[1,5]{Tadashi Kadowaki}
\affil[1]{Global R\&D Center for Business by Quantum-AI Technology, National Institute of Advanced Industrial Science and Technology, Ibaraki, Japan}
\affil[2]{(Present address) NVIDIA Corporation, 2788 San Tomas Expressway, Santa Clara, CA 95051, USA}
\affil[3]{Chemical Physics Theory Group, Department of Chemistry, University of Toronto, Toronto, Ontario, Canada}
\affil[4]{Vector Institute for Artificial Intelligence, Toronto, Ontario, Canada}
\affil[5]{DENSO CORPORATION, Tokyo, Japan} 

\date{}

\begin{document}
\maketitle

\begin{abstract}
%
%% Basic introduction
Quantum computing is entering a transformative phase with the emergence of logical quantum processors, which hold the potential to tackle complex problems beyond classical capabilities.
While significant progress has been made, applying quantum algorithms to real-world problems remains challenging.
%
%% More detailed background
%% General problem
Hybrid quantum-classical techniques have been explored to bridge this gap, but they often face limitations in expressiveness, trainability, or scalability.
%
%% Main result
In this work, we introduce conditional Generative Quantum Eigensolver (conditional-GQE), a context-aware quantum circuit generator powered by an encoder-decoder Transformer.
%
%% What the main result reveals
Focusing on combinatorial optimization, we train our generator for solving problems with up to 10 qubits, exhibiting nearly perfect performance on new problems.
By leveraging the high expressiveness and flexibility of classical generative models, along with an efficient preference-based training scheme, conditional-GQE provides a generalizable and scalable framework for quantum circuit generation.
%
%% Putting the results into a more general context
Our approach advances hybrid quantum-classical computing and contributes to accelerate the transition toward fault-tolerant quantum computing.
\end{abstract}

\begin{figure}[t]
    \centering
    \includegraphics[clip, width=0.8\columnwidth]{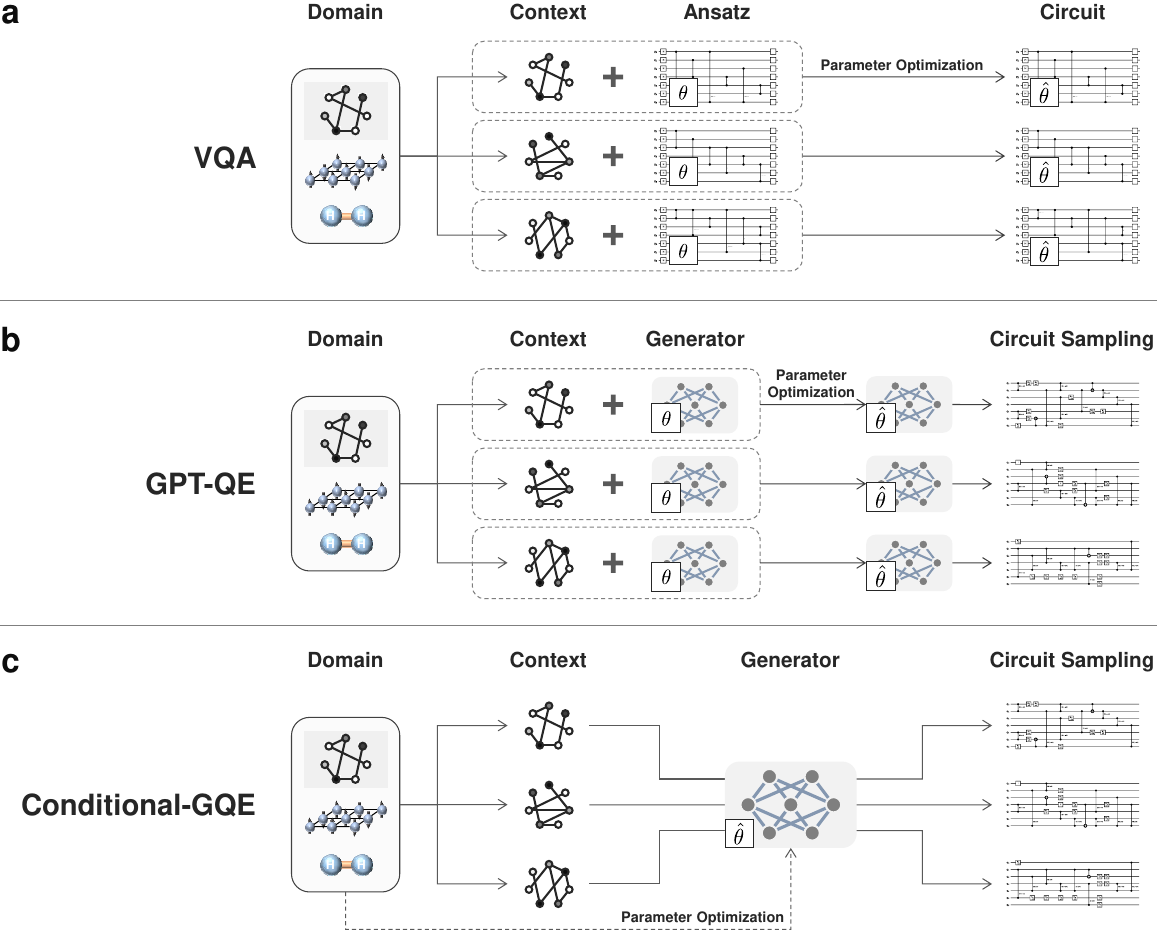}
    \caption{
        Schematic of differences between VQA, GPT-QE, and conditional-GQE.
        (a) 
        VQAs such as VQE prepare a parameterized quantum circuit, called ansatz, for each context (i.e., target problem) and optimizes the parameters to minimize the expected value of the observables.
        (b) 
        GPT-QE optimizes the parameters for each context; however, the parameters are given as weights in a classical neural network instead of being embedded in the quantum circuit.
        The final results are obtained by sampling circuits from the trained model. In the current version of the algorithm, one needs to retrain the model whenever a new problem is given. 
        (c) 
        This study develops a context-aware quantum circuit generator by using an encoder-decoder structure that enables the model to be conditioned on the problem context. 
        Once trained, the model can be used for any context in the domain and does not necessarily need to be re-trained.
    }
    \label{fig:vqa-vs-gqe}
\end{figure}

\section{Introduction}
We are at the dawn of the era of fault-tolerant quantum computation. 
Logical qubits have been demonstrated across several quantum computing architectures~\cite{ryan2022implementing,sivak2023real,Bluvstein2024,putterman2024hardware,Reichardt2024demonstration,acharya2024quantum}.
Recent advances in error-correcting codes~\cite{bravyi2024high,xu2024constant} further accelerate the shift toward early fault-tolerant systems in the relatively near future.
While these developments enable substantially more quantum operations than systems without error correction, executing fault-tolerant quantum algorithms for practically significant problems remains a distant goal. 
Thus, building hybrid quantum-classical algorithms that work with quantum devices in the early-fault-tolerant regime is a practical and reasonable focus~\cite{bharti2022noisy,alexeev2024artificial} for near-term quantum computing applications.

One widely studied methodology over the past decade is the variational quantum algorithm (VQA)~\cite{peruzzo2014variational, mcclean2016theory, cerezo2021variational, RevModPhys.94.015004}. 
Applications of VQA, such as quantum machine learning~\cite{lloyd2013quantum, Schuld03042015, biamonte2017quantum, mitarai2018quantum}, often require uploading classical data into the circuit. 
The most common strategy is to embed the data into the rotation angles of gates in a parameterized quantum circuit (PQC). 
However, this approach faces limitations in expressibility, as the Fourier components of the expectation function are constrained to specific wave numbers \cite{perez2020data,gil2020input,schuld2021effect}. 
Moreover, embedding classical knowledge or inductive bias into the PQC structure remains challenging \cite{kubler2021inductive}, despite the critical role of inductive bias in successful optimization~\cite{baxter2000model}. 
Addressing these limitations requires innovative strategies that lead to the next-generation hybrid quantum-classical computation.

This paper explores an alternative approach based on the recently proposed generative quantum eigensolver (GQE)~\cite{nakaji2024generativequantumeigensolvergqe}. 
GQE is a hybrid quantum-classical algorithm that uses a classical generative model to construct quantum circuits, where circuit components are sequentially generated from a predefined gate pool, similar to sentence generation in natural language processing.
Unlike VQAs, no parameters are embedded in the quantum circuit; all parameters are contained within a classical generative model (see Fig.~\ref{fig:vqa-vs-gqe}). 
These parameters are iteratively updated to minimize a particular objective.
In a proof-of-concept experiment~\cite{nakaji2024generativequantumeigensolvergqe}, the generative model is implemented using GPT-2~\cite{radford2019language} architecture, referred to as the generative pre-trained transformer quantum eigensolver (GPT-QE), and its effectiveness is demonstrated in the ground state search of molecular Hamiltonians.
A key feature of GQE is its ability to incorporate classical variables directly into the neural network, allowing for a non-trivial influence on the generated quantum circuits.
Additionally, inductive biases can be seamlessly integrated, much like classical convolutional neural networks in computer vision~\cite{krizhevsky2012imagenet, simonyan2014very, He_2016_CVPR} and graph neural networks in materials informatics~\cite{NIPS2015_f9be311e, kearnes2016molecular, reiser2022graph}.
While the potential of incorporating classical variables into the generative model has been previously discussed in the context of quantum chemistry~\cite{nakaji2024generativequantumeigensolvergqe}, specific methods for its implementation have not yet been detailed.

Based on the concept of GQE, this paper introduces conditional-GQE (Fig.~\ref{fig:vqa-vs-gqe}c), an input-dependent quantum circuit generation. 
To generate circuits from given inputs, we adopt an encoder–decoder Transformer architecture~\cite{vaswani2017attention}, making the model applicable across different contexts.
We apply this conditional-GQE approach to combinatorial optimization and develop a new hybrid quantum-classical method called Generative Quantum Combinatorial Optimization (GQCO).
By incorporating a graph neural network~\cite{sato2020survey} into the encoder to capture the underlying graph structure of combinatorial optimization problems, our model is trained to generate quantum circuits to solve combinatorial optimization problems with up to 10 qubits, achieving about 99\% accuracy on new test problems.
Notably, for 10-qubit problems, the trained model finds the correct solution faster than brute-force methods, simulated annealing (SA)~\cite{kirkpatrick1983optimization}, or the quantum approximate optimization algorithm (QAOA)~\cite{farhi2014quantum}.

Many of the existing works for quantum circuit design~\cite{furrutter2024quantum, PhysRevApplied.22.L041001} often rely on labeled datasets, which limits their scalability as classical simulation becomes infeasible for large circuits. 
Although some recent approaches explore reinforcement learning for circuit optimization~\cite{zhang2020topological, Preti2024hybriddiscrete}, they typically require computing intermediate quantum states to guide gate selection.
Consequently, both these supervised and reinforcement learning methods become impractical for large-scale quantum systems where classical simulation of the quantum algorithm is not feasible.
To address these challenges, we introduce a dataset-free, preference-based algorithm.
Specifically, this work uses direct preference optimization (DPO)~\cite{rafailov2024direct} to update the circuit parameters by comparing the expected values of generated circuits. 
Unlike many supervised or reinforcement learning–based methods, our DPO-based strategy does not rely on prior labeling; it only requires the final measurement results of the generated circuits, thereby substantially reducing computational overhead.

While we illustrate the potential of conditional-GQE by focusing on combinatorial optimization, our broader goal is to present a novel, scalable, and generalizable workflow for quantum circuit generation across diverse domains, which is accelerated with the help of high-performance computing systems~\cite{alexeev2024artificial}. 
This work is expected to support practical quantum computation in the early fault-tolerant era and advance quantum technology's real-world application.

\section{Results}
\subsection*{Generative quantum eigensolver (GQE)}
Large language models (LLMs) generate sequences of token indices, each corresponding to a word or subword, which, in turn, form a sentence together.
Analogously, GQE generates quantum circuits by mapping each index to a component of a quantum circuit, such as a gate or a gate combination. 
The generated sequence of indices results in a composition of quantum gates, forming a quantum circuit.

Given a fixed initial state $\ket{\phi_{\rm ini}}$, GQE uses classical machine learning to generate a quantum circuit $U$ that minimizes the expectation value $\defexpectation := \bra{\phi_{\rm ini}} U^\dagger \mathcal{O} U \ket{\phi_{\rm ini}}$ of an observable $\mathcal{O}$. 
In many quantum computing applications, observables can be expressed as the function $\mathcal{O}(\myvec{x})$ of certain variables $\myvec{x}$, such as coefficients of the Ising Hamiltonian representing combinatorial optimization problems.
However, similar to many VQAs, GPT-QE--—the original demonstration of GQE--—does not incorporate $\myvec{x}$ into the generative model but instead uses a separate model for each context, as illustrated in Fig.~\ref{fig:vqa-vs-gqe}a-b.
We believe that incorporating contextual inputs into the generative model can yield significantly different results compared to previous algorithms. 
This study presents the context-aware GQE, which aims to train a generative model with contextual inputs, generating a circuit that minimizes the energy $\expectationl{U}$ in response to a given input $x$. 
In contrast to GPT-QE, which utilizes a decoder-only Transformer, we employ a Transformer architecture that includes both an encoder and a decoder.
The details of GQE and our approach are provided in the Methods section.

In previous work by some of us \cite{cervera2021meta}, we suggested a way of training a parameterized quantum circuit $U(\myvec{\theta}, \myvec{x})$ depending on the variables $\myvec{x}$. 
In this algorithm, the variables $\myvec{x}$ are embedded into the circuit, and the parameters $\myvec{\theta}$ are optimized so that $\expectation{U(\myvec{\theta}, \myvec{x})}$ is minimized for each $\myvec{x}$. 
However, embedding classical data into a parameterized quantum circuit faces the challenge of expressibility \cite{perez2020data,gil2020input,schuld2021effect}, meaning that the functional form of $U(\myvec{\theta}, \myvec{x})$ for $\myvec{x}$ is severely restricted. 
In contrast, in GQE, we are not restricted by these expressibility issues. 
The variables $\myvec{x}$ are incorporated into the classical neural network alongside trainable parameters, and they affect non-trivially the generated quantum circuit.

\subsection*{Conditional quantum circuit generation for combinatorial optimization}
\begin{figure}[t]
    \centering
    \includegraphics[clip, width=0.8\columnwidth]{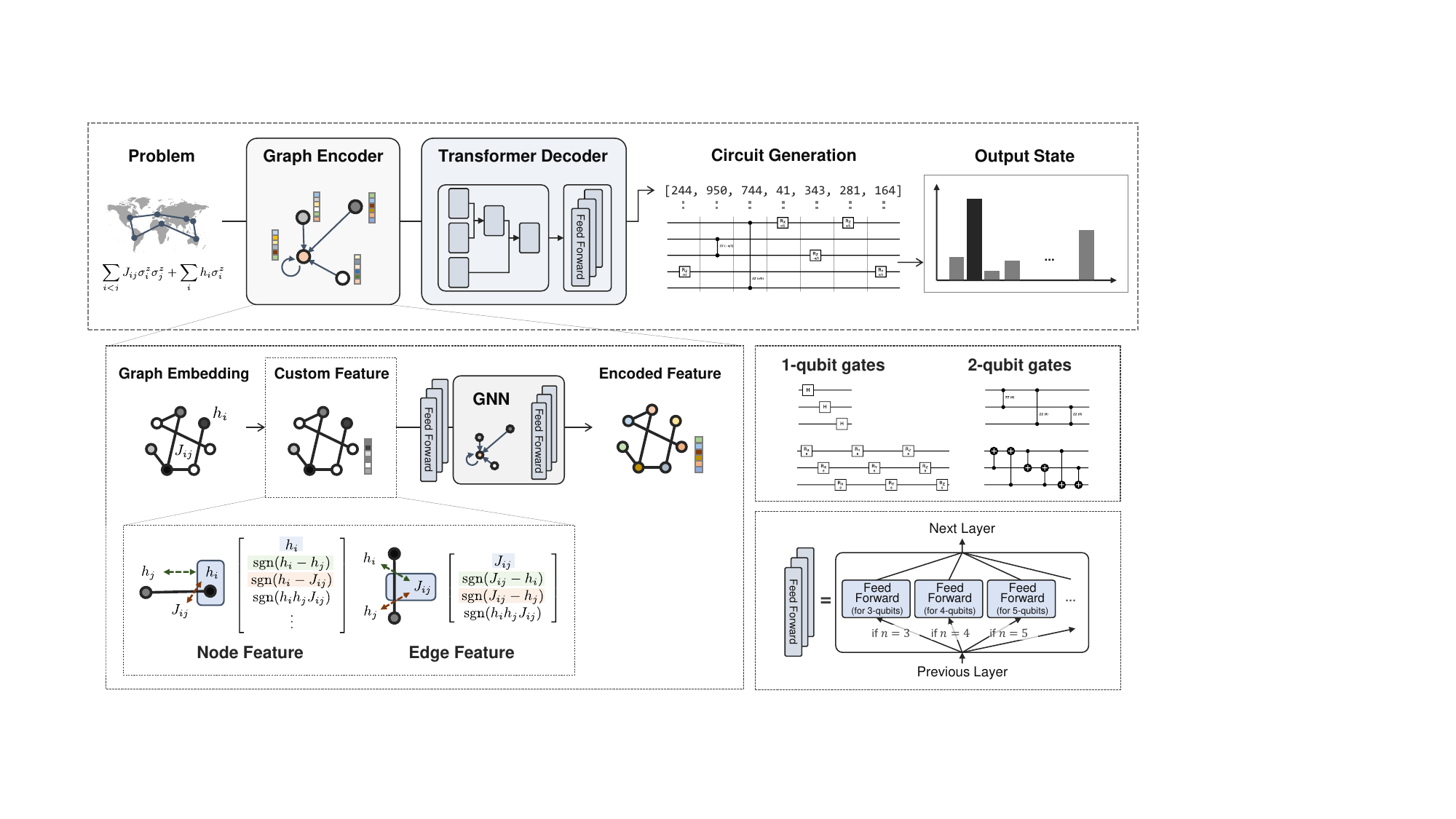}
    \caption{
        Overview of generative quantum combinatorial optimization (GQCO).
        GQCO employs an encoder-decoder Transformer architecture. 
        The target combinatorial optimization problem is represented as a graph derived from the coefficients of the corresponding Ising model. 
        Features are engineered based on domain knowledge, and an encoded feature representation is obtained using a graph neural network. 
        The encoded feature is passed to a decoder Transformer, which sequentially generates token indices and constructs sequences of 1- or 2-qubit quantum gates. 
        The mixture-of-experts (MoE) architecture is used in the model structure to improve the model expressiveness.
        The solution to the input problem is obtained from the quantum states computed by the generated circuit.
        }
    \label{fig:overview}
\end{figure}
As a very important practical application, we focus on solving combinatorial optimization problems with conditional-GQE, which we call Generative Quantum Combinatorial Optimization (GQCO).
The schematic diagram of the entire workflow is shown in Fig.~\ref{fig:overview}.

Combinatorial optimization problems can always be mapped to a corresponding Ising Hamiltonian~\cite{lucas2014ising}, which serves as an observable.
We use the Hamiltonian coefficients as input to a generator, embedding them into graph structures with feature vectors that capture domain-specific information.
A graph encoder with Transformer convolution~\cite{shi2020masked} then produces an encoded representation.
Using this, a Transformer decoder generates a sequence of token indices that defines a quantum circuit. 
The solution is identified by selecting the bit sequence corresponding to the computational basis state with the highest observation probability from the generated quantum circuit.
Fig.~\ref{fig:overview} presents the schematic of this process, with further details provided in the Methods section.

Circuit component pools must be predefined, allowing for the incorporation of domain knowledge and inductive bias. 
For example, since GPT-QE~\cite{nakaji2024generativequantumeigensolvergqe} aims to search a ground state for a given molecule, the operator pool is composed of unitary coupled-cluster singles and doubles (UCCSD) ansatz~\cite{barkoutsos2018quantum} derived from the target molecule. 
In this study, we use basic 1- and 2-qubit gates (Hadamard gate, rotation gates, and CNOT gate) and the QAOA-inspired $R_{ZZ}$ rotational gate, i.e., an Ising-ZZ coupling gate acting on two target qubits. 
The target qubit(s) of each quantum gate and, if necessary, the control qubit, are available in all configurations, and there are six possible rotation angles of $\{ \pm \frac{\pi}{3}, \pm \frac{\pi}{4}, \pm \frac{\pi}{5}\}$ for the rotation gates. 
By using basic gates rather than components suitable for many-body physics such as the UCCSD ansatz, this work aims to study whether we can train the model successfully without prior knowledge of an optimal or intuitively useful operator pool.

While a detailed description of our training strategy is provided in the Methods section, we summarize it here to highlight our scalable, broadly applicable framework. 
Scaling circuit size is critical for fault-tolerant quantum computing; however, most prior works~\cite{furrutter2024quantum, PhysRevApplied.22.L041001} rely on supervised learning methods that struggle to produce high-quality training data at large scales.
In contrast, GPT-QE employs an alternative training approach called logit matching.
This method does not require any pre-existing dataset; instead, it trains the generative model to approximate a Boltzmann distribution derived from the expectation value of a given Hamiltonian.
In this work, to further increase the probability around the preferred circuits beyond what is computed by the Boltzmann distribution, we use a preference-based strategy called direct preference optimization (DPO)~\cite{rafailov2024direct}.
DPO compares candidate circuits based on their computed costs and updates the model parameters to increase the likelihood of the most favourable circuit.
Crucially, it relies solely on expectation values from the generated circuits, eliminating the need for labelled datasets and therefore it facilitates the treatment of large-scale quantum systems.
In other words, the model is trained by exploring the space of solutions rather than relying on previously-gathered ground truth. 
To manage the diversity arising from different problem sizes, we introduce a qubit-based mixture-of-experts (MoE) architecture~\cite{jacobs1991adaptive, shazeer2017outrageously, fedus2022switch}.
This module comprises specialized model sublayers called experts, and the model switches between layers depending on the number of qubits required.
We further accelerate model training through curriculum learning~\cite{soviany2022curriculum}, starting with smaller circuits and increasing task complexity step by step, then we proceed to fine-tune each expert for the respective problem size. 
Our preference-based curriculum training with MoE modules enhances the model’s expressive power and scalability, facilitating the efficient integration of larger quantum circuits.

\subsection*{Solving combinatorial optimization via GQCO}
\begin{figure}[t]
    \centering
    \includegraphics[clip, width=0.9\columnwidth]{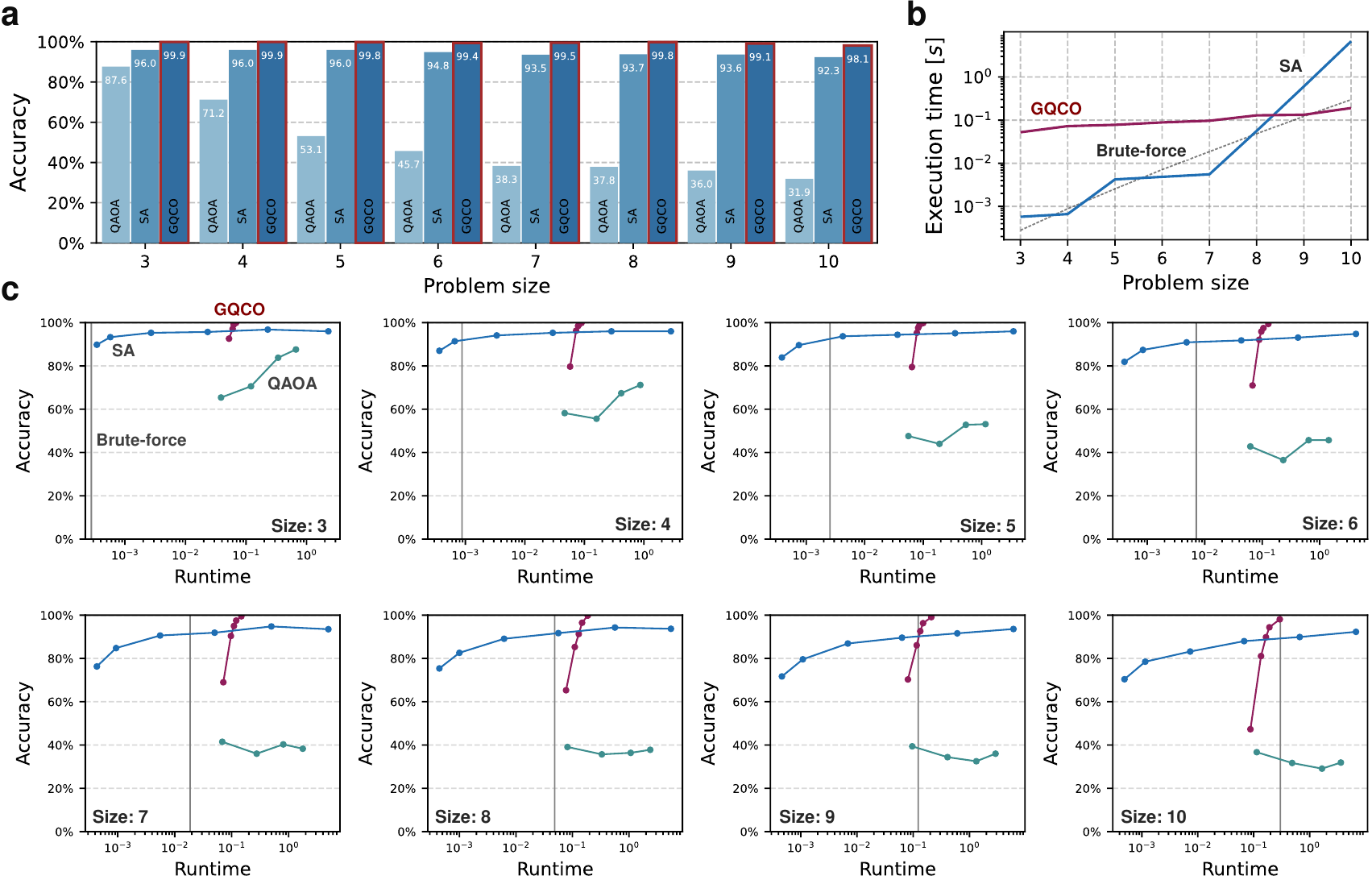}
    \caption{
        Performance evaluation of GQCO and two other solvers.
        (a) 
        Percentage of correct answers of QAOA, SA, and GQCO on 1,000 randomly generated combinatorial optimization problems (3–10 qubits).
        (b) 
        Runtime required to reach 90\% accuracy.  
        The red line represents GQCO, the blue line represents SA, and the gray dashed line represents the brute-force calculations. 
        QAOA is excluded as it did not achieve 90\% accuracy.
        (c) 
        Runtime versus accuracy across problem sizes.
        As in (b), the red lines correspond to GQCO, the blue lines to SA, and the green lines to QAOA. 
        Gray vertical lines show brute-force execution times; points to the left indicate a faster runtime than brute force.
        The points for each solver correspond to varying parameter settings: the number of sampling circuits $\{1, 5, 10, 20, 100\}$ for GQCO, the number of sweeps $\{10^2, 10^3, 10^4, 10^5, 10^6, 10^7\}$ for SA, and the number of layers $\{1, 2, 3, 4\}$ for QAOA.
    }
    \label{fig:performance}
\end{figure}
We trained a GQCO model capable of generating quantum circuits with 3 to 10 qubits.
All computations during the training were performed on classical hardware (CPUs and GPUs), and quantum calculations were conducted using a classical simulator.
Specifically, multiple NVIDIA V100 GPUs were used for the GPU computations.
Details of the training and hardware are provided in the Method section and the Supplementary Information.

Fig.~\ref{fig:performance}a compares the accuracy of GQCO with two other solvers—--simulated annealing (SA)~\cite{kirkpatrick1983optimization} and the quantum approximate optimization algorithm (QAOA)~\cite{farhi2014quantum}—---on 1,000 randomly generated combinatorial optimization problems for each problem size. 
After training the GQCO model, we selected the circuit with the lowest expected value among 100 sampled circuits for each test problem.
SA and QAOA were initialized and performed independently for each problem; in particular, in QAOA, the circuit parameters were trained from scratch for each problem and the solution was determined based on the optimized circuit.
The GQCO model consistently achieved a high accuracy of approximately 99\% across all problem sizes, whereas the other two solvers gradually decreased accuracy as the problem size increased.
Notably, QAOA failed to exceed 90\% accuracy even for a 3-qubit task, and its accuracy declined to about 30\% for a 10-qubit task.
This performance drop reflects the limited expressive power and trainability of the canonical QAOA approach.
Achieving over 90\% accuracy with QAOA would require a much deeper parameterized circuit, making, at the current time, stable training infeasible.
In contrast, GQCO addresses these limitations by leveraging the high expressive power of classical neural networks and by employing a large number of parameters on the classical side of the computation.
The performance gap observed here indicates the advantages of the generative quantum algorithm approach over variational algorithms.

A notable difference between our numerical exploration of GQCO and SA, was the runtime of the two algorithms. 
Fig.~\ref{fig:performance}b shows the time required for each method to reach 90\% accuracy.
To adjust runtime, we varied the hyperparameters---namely, the number of sampled circuits for GQCO, the number of sweeps for SA, and the number of iteration layers for QAOA.
The total runtime includes all steps, from submitting a test problem to identifying the answer.
For GQCO, this runtime encompasses both model inference and quantum simulation; for QAOA, it includes parameter optimization and quantum simulation.
SA and brute-force calculations were performed on CPUs, while the other computations, including quantum simulation, were conducted on GPUs.
The gray dashed line indicates the runtime of brute-force search, which grows exponentially with problem size; SA exhibits a similar exponential trend.
In contrast, the increase in GQCO's runtime was surprisingly restrained.
Although it took a certain amount of computation even for small problem sizes due to the need for Transformer inference, GQCO surpassed the brute-force method when the problem size exceeded 10 qubits. 
In terms of computational complexity, the brute-force method for a problem size $n$ requires a runtime on the order of $O(2^n)$.
In contrast, GQCO's complexity depends on both Transformer inference and the quantum computation of the generated circuit. 
The former depends on sequence length~\cite{vaswani2018tensor2tensor} (i.e., circuit depth) and scales on the order of $O(n^2)$, while the latter can potentially benefit from exponential speedup on quantum devices. 
In other words, GQCO can be expected to provide polynomial acceleration compared to brute-force, though GQCO does not guarantee to reach 100\% accuracy.
It is important to note that, in this performance evaluation, the quantum computations were performed using a GPU-based simulator, so any speedup that could be gained from a quantum approach would not be present in any of these results.
Nevertheless, a clear reduction in the growth rate of the runtime even for these classical simulations observed.

Fig.~\ref{fig:performance}c illustrates the detailed relationship between runtime and accuracy. 
Generally, performance improved as execution time increased. 
However, for QAOA, increasing the number of training iterations did not consistently enhance the performance of the algorithm, especially with an increasing number of qubits.
This behaviour is attributed to the training difficulties inherent in VQAs. 
In contrast, GQCO outperformed the other solvers, demonstrating greater performance gains as the execution time grew.
This advantage arises from the processing power of GPUs, which enables additional  samplings at little additional wall-clock cost, thereby boosting performance.

Because the runtime baseline depends on a device’s computational power, the problem size at which the advantage emerges may differ across devices. 
However, the difference in computational complexity is independent of the device used.

\subsection*{Error analysis of GQCO solutions}
\begin{figure}[t]
    \centering
    \includegraphics[clip, width=0.75\columnwidth]{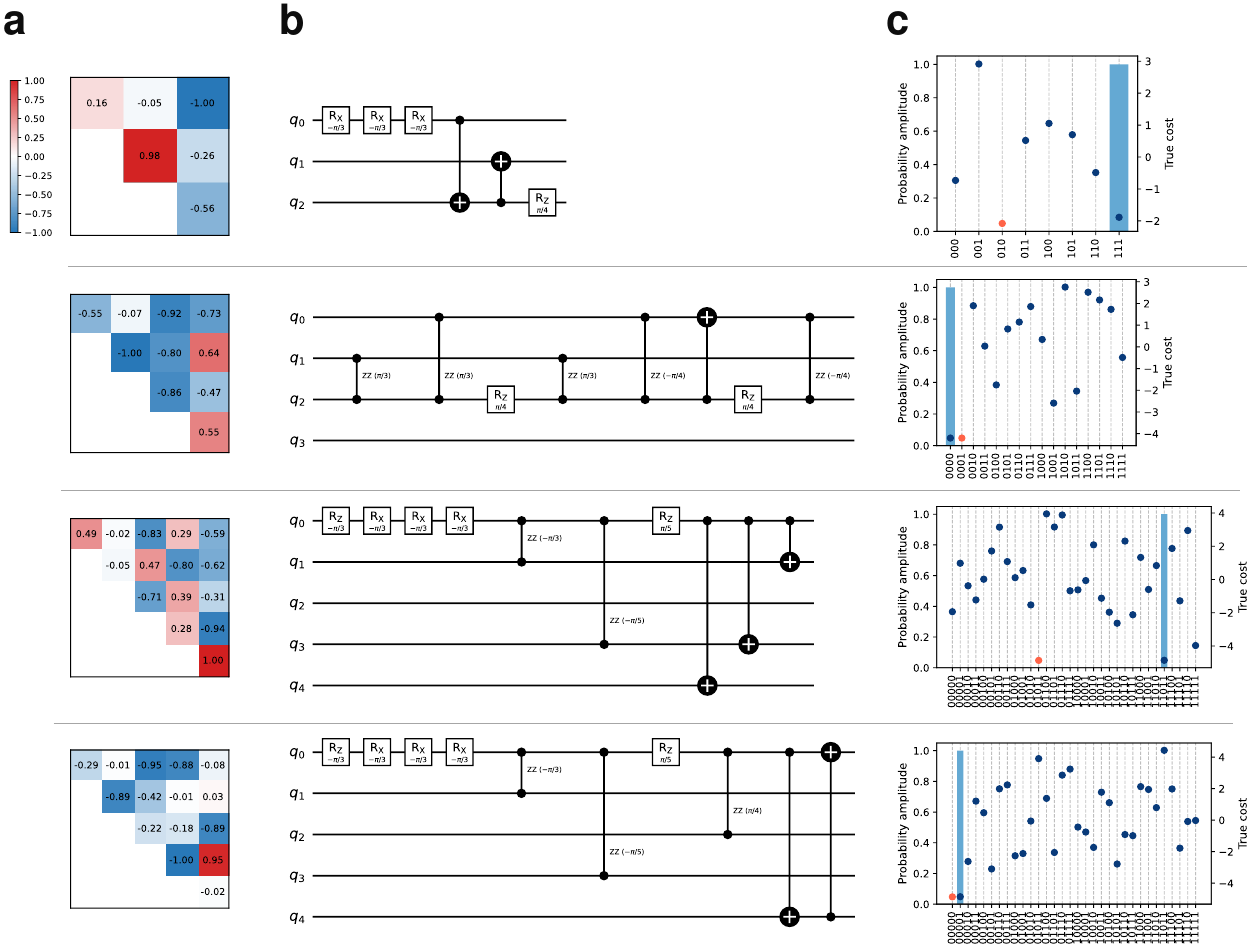}
    \caption{
        Cases where the GQCO model failed to identify the correct solution.
        (a) 
        Heat maps of the Ising Hamiltonian coefficient matrices for four incorrectly solved problems, with diagonal elements representing external fields and off-diagonal elements representing interaction terms.
        (b) 
        Quantum circuits with the lowest expected energy, selected from 100 circuits generated by the GQCO model for each combinatorial optimization problem.
        (c) 
        Corresponding quantum states obtained from these circuits.
        The histograms show observation probability for each computational basis (left y-axis) computed by state vector simulations, while the point plots show the Hamiltonian expectation value (i.e., the cost of the combinatorial optimization problem) computed in each computational basis (right y-axis). 
        The red dot for each plot corresponds to the basis with the lowest expected value, indicating the ground truth.
    }
    \label{fig:mistakes}
\end{figure}
During the performance evaluation, we identified one incorrect answer in each size-3 and size-4 problem set and two in the size-5 problem set.
Fig.~\ref{fig:mistakes} shows the corresponding coefficient matrices, generated quantum circuits, and resulting quantum states.
As mentioned above, we sampled 100 circuits for each problem, with the circuit yielding the lowest expected value among them identified as the GQCO answer.
Note that lower-cost configurations may exist outside this finite sampling; indeed, for the four problems in Fig.~\ref{fig:mistakes}, sampling 100 circuits alone did not produce a correct solution.
In each case, the second-best solution had a cost that was very close to the optimal value, causing the GQCO model to become trapped in a near-suboptimal solution.
This likely occurred because the Transformer cannot fully capture the discrete nature of combinatorial optimization, where slight fluctuations in the Hamiltonian coefficients can lead to discontinuous changes in the solution.
Increasing training time or using more precise floating-point calculations may help in reducing such errors.
Another effective approach is to increase the number of circuit samplings; indeed, the four problems in Fig.~\ref{fig:mistakes}a were correctly solved by GQCO when 1600, 300, 400, and 700 circuits were sampled, respectively.
In natural language processing, the inference scaling law~\cite{chen2024more, brown2024large, snell2024scaling} states that increasing the inference time improves the quality of model outputs.
A similar phenomenon appears to apply to quantum circuit generation as well.
However, because generative models are inherently stochastic, theoretically guaranteeing perfect accuracy for circuit generation remains challenging.

\subsection*{Characteristics of generated circuits and limitations of GQCO}
\begin{figure}[t]
    \centering
    \includegraphics[clip, width=0.95\columnwidth]{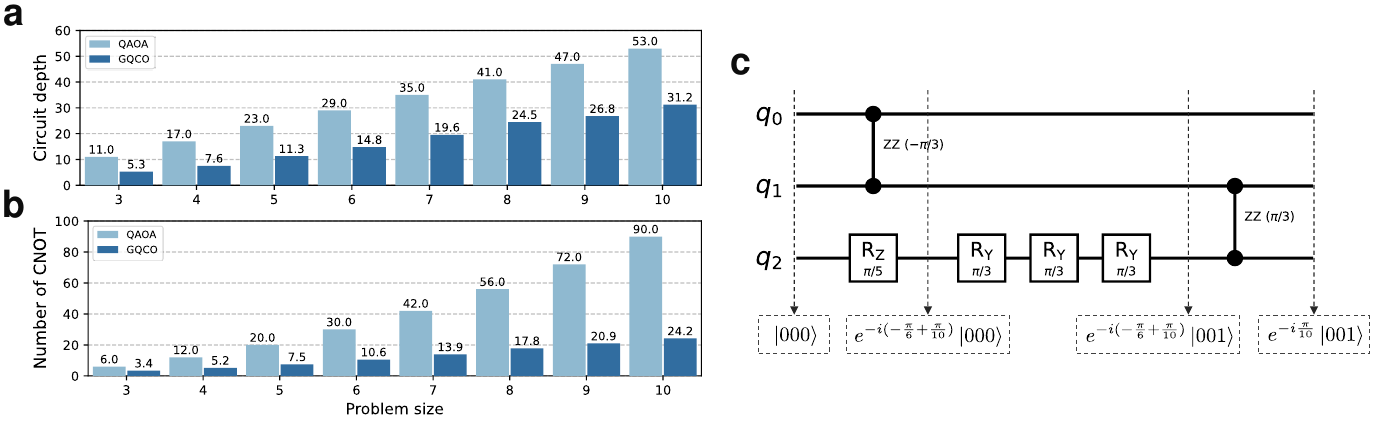}
    \caption{
        Analyses of the generated circuits.
        (a) Comparison of circuit depth and (b) number of CNOT gates between GQCO-generated and one-layer QAOA circuits for each problem size.
        (c)
        A representative example of a generated circuit.
        Six quantum gates are applied sequentially to the initial state $\ket{000}$ to obtain the final quantum state $e^{-i\frac{\pi}{10}}\ket{001}$. 
        The quantum states at two intermediate stpes are also illustrated.
    }
    \label{fig:limitation}
\end{figure}
Examining the structure of circuits generated by GQCO offers insight into how GQCO solves problems. 
Fig.~\ref{fig:limitation}a-b show the average circuit depth and the number of CNOT gates for circuits generated by the GQCO model and a single-layer QAOA circuit, respectively. 
Both values were obtained after transpiling the circuits using Qiskit~\cite{qiskit2024} with optimization level 1 (light optimization). 
Notably, the GQCO-generated circuits are shallower and include fewer CNOT gates than those produced by QAOA. 
Because the GQCO algorithm does not use the Hamiltonian directly in its circuit design, it does not require the extensive entanglement operations that QAOA does.

In this work, we did not impose an explicit restriction on the number of CNOT gates, although the maximum circuit depth for GQCO was set to twice the number of qubits.
Certainly, the cost function and model structure are flexible enough to incorporate additional constraints on circuit depth or CNOT gate count. Further constraints that lead to shallower circuits could help generate circuits more robust to noise.
Furthermore, the model can address device-related constraints.
Because many quantum devices have restricted physical connectivity~\cite{arute2019quantum, kim2023evidence}, compilation is often needed to map circuits onto the hardware. 
However, GQE-based quantum circuit generators can bypass this process by excluding gates that do not satisfy the device’s physical constraints.
This flexibility in generating hardware-efficient circuits is a key advantage of the GQE-based approach.

Fig.~\ref{fig:limitation} shows a typical 3-qubit quantum circuit generated by our GQCO model. 
In this circuit, six gates are used to transform the initial state $\ket{000}$ to state $e^{-i\frac{\pi}{10}}\ket{001}$. 
Notably, the three successive $R_Y(\pi/3)$ gates placed in the middle of the circuit are primarily responsible for obtaining the final state. 
The matrix representation of the composition of three $R_Y(\pi/3)$ gates is given by 
$
\Big[
\begin{array}{cc}
    0 & -1\\
    1 & 0
\end{array}
\Big]
$, 
which corresponds to a bit flip from $\ket{0}$ to $\ket{1}$ (or from $\ket{1}$ to $-\ket{0}$). 
The remaining three gates (one $R_Z$ gate and two $R_{ZZ}$ gates) only change the global phase for the computational basis states and have no direct effect on the final solution. 
These observations suggest that GQCO differs substantially from quantum-oriented methods such as QAOA in that the GQCO model does not acquire a quantum mechanics-based logical reasoning capability.
Instead, much like many classical machine learning models, GQCO appears to generate circuits by interpolating memorized instances.
GQCO's circuit-generation ability relies on a data-driven approach rather than any logical understanding of quantum algorithms.
However, it should be noted that combinatorial optimization problems are not inherently quantum, and the ground truth is represented by one of the computational basis state.
Because such problems can be solved with simple bit-flip circuits, it is natural that GQCO converges on generating such trivial circuits.
For more distinctly quantum tasks, such as molecular ground state searches, conditional-GQE is expected to acquire the ability to generate physically interpretable circuits.

All of the circuits generated during the performance comparison (Fig.~\ref{fig:performance}) are non-Clifford circuits and are generally expected to be difficult to simulate classically. 
However, as noted above, many of these circuits primarily perform bit flips. 
If we remove gates that affect only the global phase (e.g., the first $R_{ZZ}$ gate in Fig.~\ref{fig:limitation}c), most GQCO-generated circuits become Clifford, allowing them to be classically simulable. 
Consequently, our findings do not demonstrate a quantum advantage~\cite{arute2019quantum} or the quantum utility~\cite{kim2023evidence} of GQE-based circuit generation.
Nevertheless, even if the trained model produces circuits that can be classically simulated, non-Clifford circuits are still generated during training. 
In other words, the entire circuit space---including circuits that are computationally hard to simulate classically---must be explored to obtain the trained model, highlighting the benefits of incorporating quantum computation into the overall workflow.
Moreover, for applications beyond combinatorial optimization, solutions often involve more complex quantum states, and the circuits generated by the trained model are expected to be classically unsimulable.
Since our model can be trained without explicitly determining whether the generated circuits are classically simulable, the GQCO workflow applies equally well to problems that rely on superposition or entanglement.
However, training generators for such problems entails a more intricate cost landscape, making it challenging to train using simple gate pools or vanilla DPO loss, as done in this study.
Future research focusing on more carefully designed workflows would therefore be promising.

\subsection*{Solving combinatorial optimization using a quantum device}
\begin{figure}[t]
    \centering
    \includegraphics[clip, width=0.9\columnwidth]{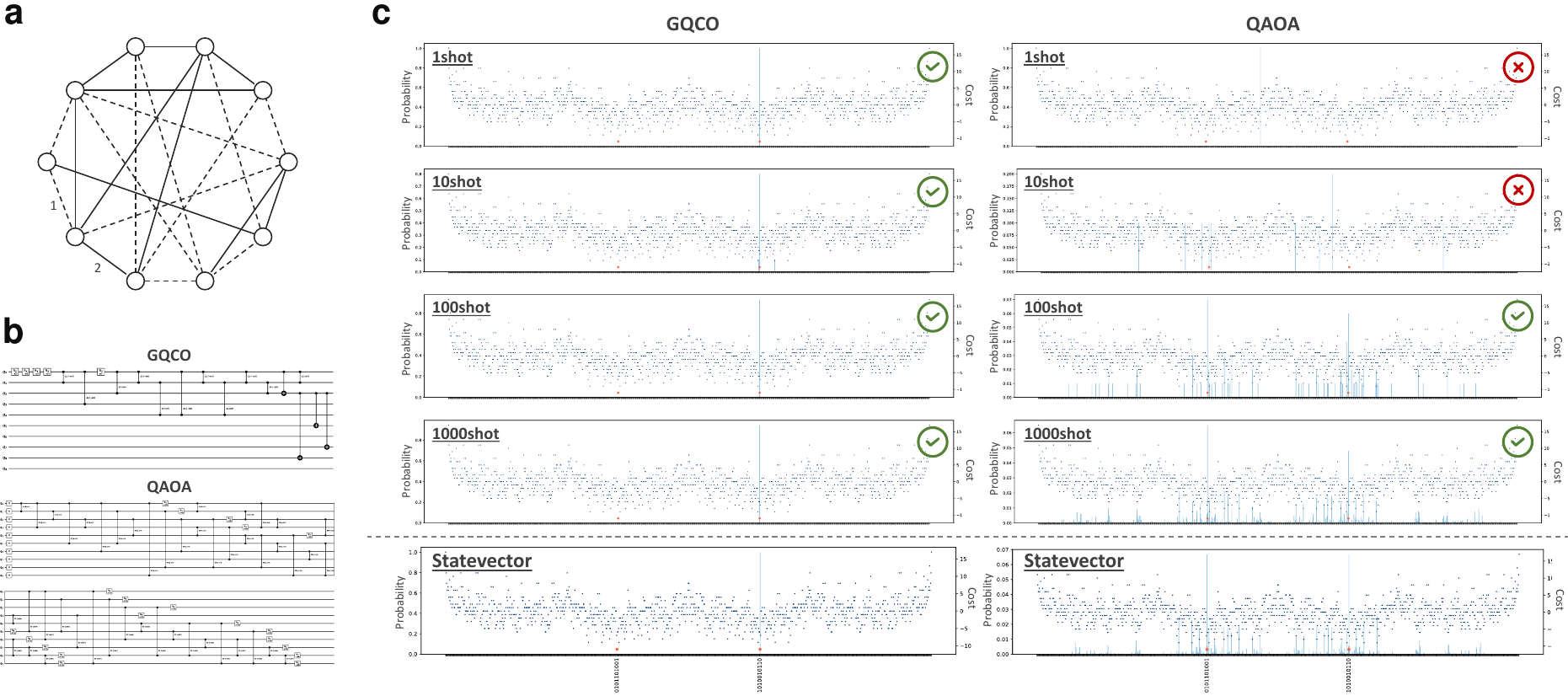}
    \caption{
        Results on the real quantum processor.
        (a)
        Target max-cut problem with 10 variables. 
        The edge weights are represented by line styles: dashed lines indicate a weight of 1, and solid lines indicate a weight of 2.
        (b) 
        Quantum circuits generated by GQCO and a two-layer QAOA circuit for the target problem.
        (c) 
        Sampling results on the real quantum device (IonQ Aria) for each of the circuits. 
        Results for 1, 10, 100, and 1,000 shots, as well as the state vectors computed by the simulator, are included. 
        The histograms and plots are interpreted in the same manner as in Fig.~\ref{fig:mistakes}c.
        Each plot is marked to indicate whether it leads to the correct solution.
        The enlarged figures are available in the Supplementary Information.
    }
    \label{fig:real}
\end{figure}
At the end of the performance analyses of GQCO, we examined its behavior on a physical quantum device.
The target problem was the 10-variable max-cut problem illustrated in Fig.~\ref{fig:real}a.
For comparison, we used a two-layer QAOA circuit whose parameters were optimized with a classical simulator.
The resulting circuits (Fig.~\ref{fig:real}b) were then executed on the IonQ Aria quantum processor.
Fig.~\ref{fig:real}c presents the sampling results for varying numbers of shots alongside the state vector computed by the classical simulator.

A key characteristic of GQCO-generated circuits is that resulting quantum state exhibits a distinct observation probability peak at a single computational basis state.
In contrast, because QAOA discretely approximates time evolution from a uniform superposition, its resulting quantum state is more complex and less likely to yield a clear peak, particularly when the circuit depth is limited. 
Consequently, QAOA required more than 100 shots to identify the correct answer in this study, whereas GQCO was able to find it with just a single shot.
This disparity in the number of required shots is expected to grow as the number of qubits increase.

Another notable aspect of GQCO becomes evident in cases where the ground state is degenerate.
In principle, our GQCO model cannot account for degeneracy because the training process relies solely on circuit sampling, focusing on identifying a solution without considering the underlying quantum mechanics of the input Ising Hamiltonian.
In a max-cut problem, the system is inherently degenerate, particularly the target problem in this section is doubly degenerate. 
As illustrated in Fig.~\ref{fig:real}c, while GQCO identified only one of the two solution candidates, QAOA exhibited observation probability peaks for both degenerated ground states. 
Originating in adiabatic quantum computation, QAOA is theoretically expected to yield a non-trivial probability distribution over degenerate ground states.
GQCO’s inability to capture degeneracy will require future work on model architectures and training approaches that incorporate quantum mechanical principles.

\section{Discussion}
We developed a novel quantum-classical hybrid technique for context-aware quantum circuit generation and then applied it to combinatorial optimization problems.
Our approach, which we have named conditional-GQE, extends GQE by integrating contextual encoding, and employs the cutting-edge methodologies such as DPO-based training and qubit-based curriculum learning to yield a scalable workflow. 
This strategy enabled us to successfully build the quantum circuit generator for combinatorial optimization, a high-performance solver that outperforms conventional solvers for problems with up to 10 variables. 
Although this work is still a prototype, the results suggest the potential for more practical, larger-scale implementations toward foundation models of combinatorial optimization.
Moreover, we highlight the capacity of classical neural networks to generate flexible, high-quality quantum circuits, paving the way for advanced quantum-classical hybrid technologies.

The conceptual workflow described in this study can be extended beyond combinatorial optimization to any problem formulated as observable expectation value minimization.
For example, in molecular ground-state searches, representing molecular structures as graphs allows direct adoption of our graph-based encoding.
By replacing the encoder, the GQE-based approach also generalizes to quantum machine learning and partial differential equation solvers.
We thus view GQE-based quantum circuit generation as a next step following VQAs.

However, several limitations remain as open problems.
A major obstacle is the significant classical computational resources required to achieve scalability. 
While our findings indicate the computational advantage over classical solvers, fully realizing this advantage demands extensive classical training beforehand.
Overcoming these challenges will likely involve designing an encoder architecture guided by domain knowledge, refining training strategies (including developing pre-training methods), and using the high-performance computing resources \cite{alexeev2024artificial}. 
Additionally, careful gate pool design will play a crucial role.
Machine learning-based approaches for identifying suitable gates or gate representation learning may offer a promising direction.

This research provides a novel pathway for quantum computation by leveraging large-scale machine learning models. 
It underscores the growing role of AI in the advancement of next-generation quantum computing research activities. 
We believe that our work will serve as a catalyst for accelerating the development of quantum applications across diverse domains and facilitating the democratization of quantum technology.

\section{Methods}
\subsection*{GQE, GPT-QE, and conditional-GQE}
Let $\myvec{t} = \{t_1, \ldots, t_N\}$ be a generated token sequence of length $N$, where each token index $t_k$ is an integer satisfying $1 \leq t_k \leq V$, with $V$ being the vocabulary size.
Each token index $t_k$ corresponds to a quantum circuit component $U_k$ selected from a predefined operator pool $\{U_\ell\}_{\ell=1}^V$. 
These components collectively form a quantum circuit $U = U_{t_N} \cdots U_{t_1}$.
Let $p_{\myvec{\theta}}(U)$ denote the generative model of quantum circuits, where $p_{\myvec{\theta}}(U)$ is a probability distribution over unitary operators $U$, and $\myvec{\theta}$ is the set of optimizable parameters.
In GQE, the parameters $\myvec{\theta}$ are iteratively optimized so that circuits sampled from $p_{\myvec{\theta}}(U)$ are more likely to minimize the expectation value of an observable:
\begin{align*}
	\langle \mathcal{O}\rangle_U
	:= \bra{\phi_{\rm ini}} U^\dagger \mathcal{O} U \ket{\phi_{\rm ini}},
\end{align*}
where $\mathcal{O}$ is an observable and $\ket{\phi_{\rm ini}}$ is a fixed input state. In particular, for an $n$ qubit system, we use $\ket{\phi_{\rm ini}} = \ket{0}^{\otimes n}$.
The quantum computation is involved in the estimation of $\defexpectation$.
Notably, unlike in VQAs, all optimizable parameters are embedded in the classical generative model $p_{\myvec{\theta}}(U)$ rather than in the quantum circuit itself (see Fig.~\ref{fig:vqa-vs-gqe}b).

As discussed in the Result section, the observable can be expressed as a function of certain variables $\myvec{x}$. 
Let us denote such an observable as $\mathcal{O}(\myvec{x})$.
The quantum circuit $U$ that minimizes $\defexpectationl$ also depends on the variable $\myvec{x}$.
In original GQE approach (including GPT-QE), parameters are set and optimized for each specific target problem, much like in VQAs.
More precisely, GPT-QE aims to obtain a decoder-only Transformer $p_{\myvec{\theta}^*(\myvec{x})}(U)$ for each $\myvec{x}$, where $\theta^*(x)$ is the solution for the following minimization problem:
\begin{align}
\label{eq:obj-gptqe}
    \myvec{\theta}^*(x) = \argmin{\myvec{\theta}} \underset{U \sim p_{\myvec{\theta}}(U)}{\mathbb{E}} \langle \mathcal{O}(\myvec{x})\rangle_U,
\end{align}
where $\mathbb{E}_{X \sim p(X)}[f(X)]$ denotes the expectation value of $f(X)$ with respect to the random variable $X$ over the sample space $\Omega$, where $X$ is drawn from the distribution $p(X)$, i.e., 
$
    \mathbb{E}_{X \sim p(X)}[f(X)] := \int_{\Omega} f(X)p(X) \mathrm{d}X
$.

By utilizing $\myvec{x}$ as context (i.e., input), conditional-GQE aims to train a generative model $p_{\myvec{\theta}}(U|\myvec{x})$ that generates circuits minimizing $\langle \mathcal{O}(\myvec{x}) \rangle_U$.
The function $p_{\myvec{\theta}}(U|\myvec{x})$ provides the conditional probability of generating the unitary operator $U$ when the input $\myvec{x}$ is given.
In Transformer-based generative models, the probability $p_\theta(U|x)$ is expressed as follows:
\begin{align*}
    p_\theta(U|x) = \prod_{i=1}^N p_\theta(U_{t_i} | U_{t_0}, \dots ,U_{t_{i-1}}, x) \propto \prod_{i=1}^N \exp \bigg\{ \frac{z_i(U_{t_0}, \dots ,U_{t_{i-1}}, x ; \theta)}{T} \bigg\},
\end{align*}
where $t_0$ is the start-token index, chosen such that $U_{t_0} = I^{\otimes n}$ in this work.
$z_i$ denotes the logit for $i$-th token, that is, the corresponding output from the model before applying the sigmoid function.
$T$ is the sampling temperature; in this work, we set $T=1.0$ for training and $T=2.0$ for evaluation, thereby enhancing randomness in the evaluation phase.
Then, we can realize a generative model $p_{\myvec{\theta^*}}(U | x)$ through the following optimization:
\begin{align}
\label{eq:obj-gqe}
    \myvec{\theta}^* = \argmin{\myvec{\theta}} \underset{x \sim p(x)}{\mathbb{E}} \ \underset{U \sim p_{\myvec{\theta}}(U|x)}{\mathbb{E}} \langle \mathcal{O}(\myvec{x})\rangle_U,
\end{align}
where $p(\myvec{x})$ denotes the probability distribution of inputs $\myvec{x}$ in the target domain.

Solving the optimization problems in Eq.~\eqref{eq:obj-gptqe} or Eq.~\eqref{eq:obj-gqe} is challenging and thus requires surrogate objective functions. 
GPT-QE employs a logit-matching approach, whereas this study utilizes DPO~\cite{rafailov2024direct} loss. 
Further details of DPO are provided in the subsequent section.

\subsection*{Outline of the GQCO model architecture}

The GQCO model proposed in this study employs an encoder-decoder Transformer architecture~\cite{vaswani2017attention} with GELU activation~\cite{hendrycks2016gaussian}.
Its architectural diagram is provided in the Supplementary Information. 
Both encoder and decoder consist of 12 repeated modules, each incorporating graph convolution~\cite{shi2020masked} and a multi-head attention Transformer. 
The encoder components are detailed in subsequent sections.
The intermediate representations have dimension 256, and each Transformer layer has 8 attention heads.
In total, the model has approximately 256 million parameters, with 127 million in the encoder and 129 million in the decoder.
This parameter count is comparable to T5~\cite{JMLR:v21:20-074}, an early encoder-decoder LLM with 246 million parameters, and the decoder alone is similar in scale to GPT-1~\cite{radford2018improving}, which is a decoder-only Transformer model and has 117 million parameters.
Our gate pool supports the generation of quantum circuits with up to 20 qubits and offers 1,901 gate candidates, including an identity gate, although we have only proceeded to the 10-qubit scale.
During generation, gates unnecessary for the given problem size are masked.
The length of the token index sequence, which corresponds to the number of generated gates, is set to a maximum of $2n$, where $n$ is the number of qubits used.
Note that, when four or more gates have been generated, and an end-token index $t_{\mathrm{end}}=t_0$ is produced, the generation process is terminated before reaching the maximum length.

\subsection*{Embedding combinatorial optimization problems as graphs}

Any combinatorial optimization problem can be bijectively mapped to an Ising Hamiltonian~\cite{lucas2014ising}: 
\begin{align*}
\mathcal{H} = \sum_{i < j} J_{ij} \sigma_i^z \sigma_j^z + \sum_{i} h_i \sigma_i^z,
\end{align*}
whose ground states correspond to the solutions of the problem. 
The coefficients of the Ising Hamiltonian---the external magnetic field $h_i$ and the interaction coefficient $J_{ij}$---are used as inputs (or contexts) to the encoder of the model. 
We map these coefficients into a graph representation, considering $h_i$ as the weight of node $i$ and $J_{ij}$ as the weight of the edge between nodes $i$ and $j$.

The feature vector is then constructed using the following three elements: (1) the weights themselves, (2) the sign of the magnitude relationships between the weights of adjacent nodes or edges, and (3) the sign of the product of the weights of adjacent nodes or edges (see Fig.~\ref{fig:overview}). 
More formally, the node feature $v_i$ and edge feature $e_{ij}$ are computed as follows:
\begin{gather*}
    v_i = \left[ 
        \begin{array}{c}
            v_i^{(1)} \\
            v_i^{(2)} \\
            v_i^{(3)} \\
            v_i^{(4)}
        \end{array}
    \right]
    ,
    e_{ij} = \left[
        \begin{array}{c}
            \sgn{J_{ij}} \\
            \sgn{J_{ij} - h_i} \\
            \sgn{J_{ij} - h_j} \\
            \sgn{h_i h_j J_{ij}}
        \end{array}
    \right]
    , \\
    v_i^{(1)} = h_i
    ,
    v_i^{(2)} = \left[
        \begin{array}{c}
            \sgn{h_i - h_{j_1}} \\
            \sgn{h_i - h_{j_2}} \\
            \vdots \\
            \sgn{h_i - h_{j_k}}
        \end{array}
    \right]
    ,
    v_i^{(3)} = \left[
        \begin{array}{c}
            \sgn{h_i - J_{ij_1}} \\
            \sgn{h_i - J_{ij_2}} \\
            \vdots \\
            \sgn{h_i - J_{ij_k}}
        \end{array}
    \right]
    ,
    v_i^{(4)} = \left[
        \begin{array}{c}
            \sgn{h_i h_{j_1} J_{ij_1}} \\
            \sgn{h_i h_{j_2} J_{ij_2}} \\
            \vdots \\
            \sgn{h_i h_{j_k} J_{ij_k}}
        \end{array}
    \right]
    ,
\end{gather*}
where $\sgn{\cdot}$ is the sign function and $\{j_\ell\}_{\ell=1}^k \big(=: \mathcal{N}(i) \big)$ denote the index set of nodes connected to node $i$.
These handcrafted features serve to incorporate domain knowledge of the Ising model into our model; specifically, the facts that spin-spin interactions with large coefficients or strong external magnetic fields have a significant impact on the spin configuration of the system, and that frustration---the absence of a spin configuration that simultaneously minimizes all interaction energies---generates a complex energy landscape.

\subsection*{Encoder with Graph Transformer convolution}
The embedded graphs are converted into encoded representations by alternately applying Graph Transformer convolutional layers~\cite{shi2020masked} and feed-forward layers. 
Specifically, local message passing and feature transformation are performed according to the following equations:
\begin{align*}
    \myvec{v}_{i}' &= \mathrm{LayerNorm} \bigg( W_1 \myvec{v}_i + \sum_{j \in \mathcal{N}(i)} \alpha_{ij} (W_2 \myvec{v}_j + W_3 \myvec{e}_{ij} ) \bigg), \\
    \alpha_{ij} &= \mathrm{softmax} \bigg( \frac{(W_4 \myvec{v}_i)^\top (W_5 \myvec{v}_j + W_6 \myvec{e}_{ij}) }{\sqrt{d}} \bigg), \\
    \myvec{v}_{i}'' &= \mathrm{LayerNorm} \big (\myvec{v}_{i}' + W_8 \ \mathrm{GELU} (W_7 \myvec{v}_{i}') \big),
\end{align*}
where the matrices $W_k (k = 1, \dots, 8)$ are trainable weight, and $d$ is the dimension of the intermediate representations (i.e. $d=256$).
$\mathrm{LayerNorm}(\cdot)$  is layer normalization~\cite{lei2016layer}, $\mathrm{softmax}(\cdot)$ denotes the softmax function, and $\mathrm{GELU}(\cdot)$ refers the GELU activation function~\cite{hendrycks2016gaussian} applied element-wise.
This base module is repeated 12 times, and the resulting node features, indexed by node order, serve as input to the decoder.
Although using 12 iterations risks issues of graph neural networks like over-smoothing~\cite{10.5555/3504035.3504468}, over-squashing~\cite{topping2022understanding}, and graph bottlenecks~\cite{alon2021on}, we set the iteration count to 12 to align the model scale with that of well-known language models such as T5 and GPT-1.

\begin{figure}[t]
    \centering
    \includegraphics[clip, width=0.9\columnwidth]{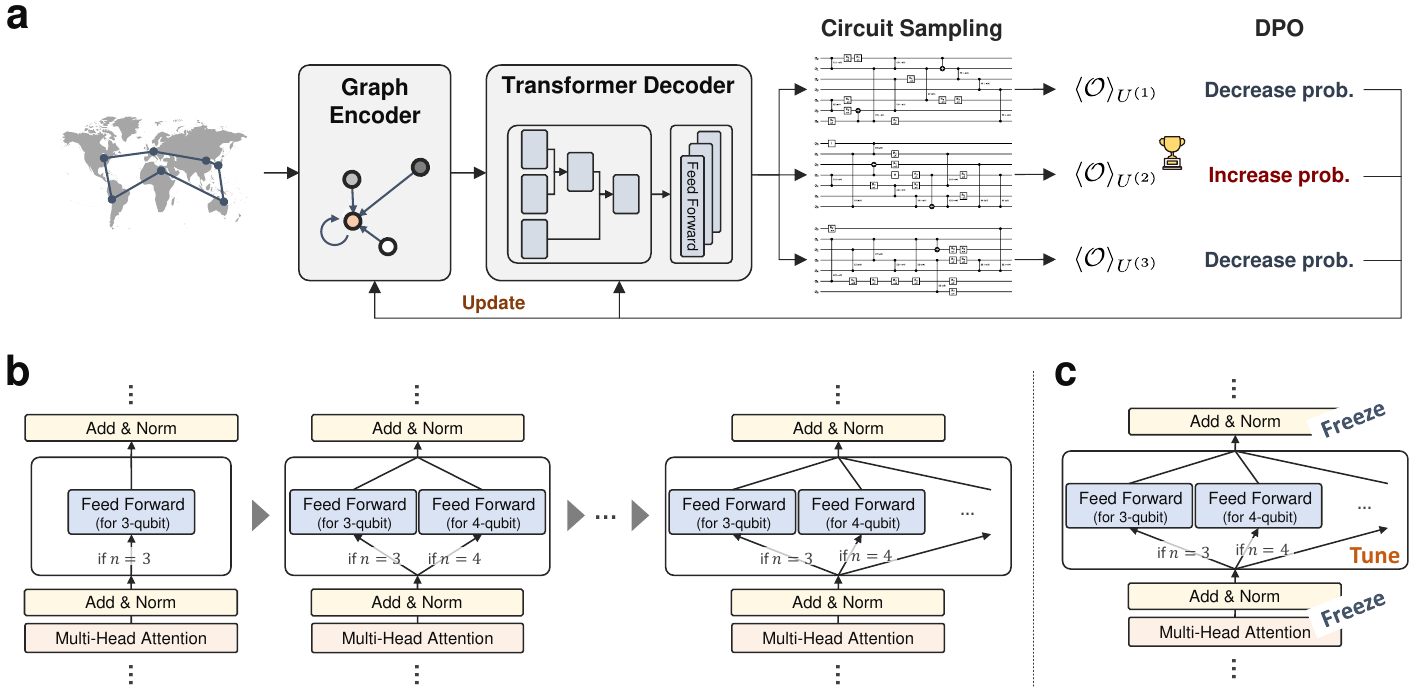}
    \caption{
        Training strategy for GQCO.
        (a) 
        Training iterations of direct preference optimization (DPO). 
        For a randomly generated input problem, multiple quantum circuits are sampled, and the expected value of the Hamiltonian is computed for each one. 
        The model parameters of the encoder and decoder are updated to increase the probability of generating the circuit with the lowest energy value while decreasing the probabilities of generating the other circuits.
        (b) 
        Curriculum learning based on the number of qubits. 
        We start training by generating circuits with three qubits and gradually increase the complexity of the task by increasing the number of qubits.
        (c) 
        Expert tuning. 
        The weights of the shared layers, such as the attention layers, are fixed, and the expert layers are fine-tuned. 
    }
    \label{fig:training}
\end{figure}

\subsection*{Direct preference optimization}
Traditionally, supervised learning with labeled dataset has been widely used in machine learning applications to quantum information processing. 
Examples include quantum circuit compilation~\cite{zhang2020topological, moro2021quantum, Preti2024hybriddiscrete}, ansatz generation for VQAs~\cite{zhang2021neural, he2023gnn}, and diffusion-based quantum circuit generation~\cite{furrutter2024quantum}. 
However, this approach faces some significant challenges.
The most notable issue is the scalability. 
Preparing training data using classical computation quickly becomes infeasible for large-scale circuits exceeding 50 qubits.
Furthermore, for complex tasks, it is difficult to prepare ground-truth circuits, or the circuit structure may not be uniquely defined. 
Consequently, preparing high-quality training datasets remains problematic.

For these reasons, we employ a preference-based approach using direct preference optimization (DPO)~\cite{rafailov2024direct}. 
DPO is a training strategy derived in the field of reinforcement learning from human feedback (RLHF)~\cite{ziegler2019finetuning, ouyang2022training}, used to fine-tune LLMs for generating preferred outputs.
In this approach, multiple outputs are sampled and parameters are updated to increase the likelihood of preferred outputs while decreasing that of less preferred ones.
Typically, in LLMs, human evaluators determine the preference of outputs.
In our study, we assess the preferrability of circuits using the computed expectation values of the Hamiltonian.

Fig.~\ref{fig:training}a shows the schematics of our DPO-based training process.
The expected DPO loss function used in this work is defined by:
\begin{align*}
    \mathcal{L}_{\mathrm{DPO}} (\theta) 
    &:= 
    \underset{x}{\mathbb{E}} \underset{w \succ \ell}{\mathbb{E}} L(w, \ell, \myvec{x} \,; \theta), \\
    L(w, \ell, \myvec{x} \,; \theta) &:= 
    \log 
    \bigg\{ 
        1 + \exp 
        \bigg\{ 
            -\beta 
            \bigg( 
                  \log \frac{p_{\myvec{\theta}}( U^{(w   )} | \myvec{x})}{\pi_{\mathrm{ref}}(U^{(w   )} | \myvec{x})} 
                - \log \frac{p_{\myvec{\theta}}( U^{(\ell)} | \myvec{x})}{\pi_{\mathrm{ref}}(U^{(\ell)} | \myvec{x})} 
            \bigg) 
        \bigg\} 
    \bigg\},
\end{align*}
where $\{ U^{(k)} \}_{k=1}^M$ is the set of sampled circuits, and $w \succ \ell$ indicates that $\expectationl{U^{(w)}} < \expectationl{U^{(\ell)}}$, i.e., the circuit $U^{(w)}$ is preferred over the circuit $U^{(\ell)}$.
$\pi_{\mathrm{ref}}(\cdot | \myvec{x})$ is a reference probability and serves as the baseline for the optimization during the DPO, and $\beta$ is a hyperparameter controlling the influence of $\pi_{\mathrm{ref}}$.
In this work, we use $\pi_{\mathrm{ref}}(U|\myvec{x}) \propto \exp \{ -\expectationl{U} \}$.
Since the negative log-sigmoid function $\log(1+\exp(-x))$ is monotonically decreasing, the function $L$ is minimized when $p_{\myvec{\theta}}( U^{(w)} | \myvec{x})$ is maximized and $p_{\myvec{\theta}}( U^{(\ell)} | \myvec{x})$ is minimized.
In other words, this loss function is designed to increase the generation probability of preferred circuits while decrease the probability of non-preferred circuits.

Ideally, the function $L$ should be computed for all possible pairs of the $M$ sampled circuits, totaling $M(M-1)/2$ combinations. 
However, to reduce computational overhead, we employ the best-vs-others empirical loss, defined as follows:
\begin{align*}
    \hat{\mathcal{L}}_{\mathrm{DPO}}^{\mathrm{BvO}} (\theta)
    := \frac{1}{|\mathcal{D}_x|} \sum_{x \in \mathcal{D}_x} \frac{1}{M-1} \sum_{\ell} L(w_{\mathrm{best}}, \ell, \myvec{x} \,; \theta),
\end{align*}
where $w_\mathrm{best}$ represents the index of the circuit with the smallest expectation value for each input $x$, and $\mathcal{D}_x$ is an input dataset with size $|\mathcal{D}_x|$.

However, if all $M$ sampled circuits are identical, the gradient of the loss will be zero regardless of the magnitude of the expected values, preventing the model from being trained effectively. 
To mitigate this issue, we employ contrastive preference optimization (CPO)~\cite{DBLP:conf/icml/XuSCTSDM024}, an improved version of DPO. 
In addition to the DPO loss, CPO introduces a negative log-likelihood term on the probability of generating the most preferred output. 
In summary, the loss function used in this work is given by:
\begin{align}
    \label{eq:cpo}
    \hat{\mathcal{L}}_{\mathrm{CPO}}^{\mathrm{BvO}} (\theta)
    := \frac{1}{|\mathcal{D}_x|} \sum_{x \in \mathcal{D}_x} \Bigg[ \frac{1}{M-1} \sum_{\ell} L(w_{\mathrm{best}}, \ell, \myvec{x} \,; \theta)
    - p_{\myvec{\theta}}( U^{(w_{\mathrm{best}})} | \myvec{x}) \Bigg].
\end{align}
The hyperparameter $\beta$ is set to 0.1, and the number of samplings $M$ is adjusted according to the number of qubits to maximize the utilization of computational memory. 
Model training proceeds by generating an input $x$ uniformly at random as a coefficient of the Ising Hamiltonian (i.e., $|\mathcal{D}_x| = 1$), computing the gradient based on the loss function Eq.~\eqref{eq:cpo}, and iteratively updating the parameters. 
See the Supplementary Information for detailed training settings, including the values of $M$.

\subsection*{Qubit-based mixture-of-experts}
Depending on the number of variables in a combinatorial optimization problem, quantum computation requires a corresponding number of qubits. 
To effectively handle the diversity of tasks resulting from variations in problem size, we employ the mixture-of-experts (MoE) architecture~\cite{jacobs1991adaptive, shazeer2017outrageously, fedus2022switch}, which is commonly used in LLMs.
As illustrated in Fig.~\ref{fig:overview} and Fig.~\ref{fig:training}b-c, each feed-forward module in the model is partitioned into specialized submodules, referred to as "experts." 
The gating mechanism dynamically selects layers based on the number of qubits, forming what we term a qubit-based MoE.
This design balances the need for diverse model representations while limiting the growth of model parameters.

\subsection*{Curriculum learning}
The qubit-based MoE enhances model scalability.
By incorporating additional experts and restarting training, circuit generators can be trained for varying numbers of qubits without the need to retrain the entire model from scratch.
Leveraging this capability, we adopt curriculum learning~\cite{soviany2022curriculum}, a method that incrementally increases task complexity by starting with simpler problems and gradually progressing to more challenging ones.

Training begins with randomly generated combinatorial optimization problems involving 3 qubits.
Performance is monitored regularly, and training continues until the model achieves an accuracy exceeding 90\% on randomly generated test problems.
Once this threshold is met, size-4 optimization problems are introduced as training candidates, along with the integration of a new expert module within the MoE layers.
Then, performance is continuously monitored, and the maximum problem size in the training dataset is gradually increased.

Even when the maximum problem size is $n_{\mathrm{max}}$, problems involving fewer qubits ($< n_{\mathrm{max}}$) are still generated as part of the training data.
The probability of generating a problem of size $n$ when the current maximum size is $n_{\mathrm{max}}$ is defined as:
\begin{align}
\label{eq:size-probability}
    p(n \mid n_{\mathrm{max}}) =
    \begin{cases} 
        0 & \text{if } n \leq n_{\mathrm{max}}< 3 \\
        1 & \text{if } n = n_{\mathrm{max}} = 3, \\
        \frac{0.5}{n_{\mathrm{max}} - 3} & \text{if }  3 \leq n < n_{\mathrm{max}}, \\
        0.5 & \text{if } 3 < n = n_{\mathrm{max}}.
    \end{cases}
\end{align}
In brief, when the number of qubits is larger than three, the probability of generating the largest problem size is $0.5$, and the remaining $0.5$ probability is equally divided among all smaller sizes.
Notably, the probability of generating previously trained problem sizes is not set to zero. 
This strategy mitigates catastrophic forgetting~\cite{kirkpatrick2017overcoming}--—a phenomenon in which performance on previously learned tasks declines rapidly in continuous learning and online learning.
Without such adjustments, continuously updating training data could lead to the model forgetting previously acquired knowledge.
A small amount of instances for smaller problem sizes helps to maintain consistent performance across all problem sizes.

\subsection*{Hardware configuration}
Multiple compute nodes, each consisting of four NVIDIA V100 GPUs, were used, and the model was trained using a distributed data parallel (DDP) strategy~\cite{10.14778/3415478.3415530}.
The quantum circuit simulations were performed on a CPU (Intel Xeon Gold 6148 processor) environment using Qiskit~\cite{qiskit2024}. 
The number of GPU nodes used in training and the epochs differ for each stage of the curriculum learning. 
The details are summarized in the Supplementary Information.

For performance evaluation, the model inferences and quantum circuit simulations were both performed on an NVIDIA RTX A6000 GPU. 
All other classical computations were performed on an Intel Core i9-14900K CPU. 
For the real quantum device, we used IonQ Aria via Amazon Braket.

\section{Data availability}
All training datasets were randomly generated during training and were therefore not stored. 
Instead, the random seeds are fixed to ensure reproducibility. 
These seeds, along with the dataset and the data generation code used for performance evaluation, will be available on GitHub. %are available at \msh{[GitHub URL]}.

\section{Code availability}
The code used for training and evaluation, as well as the trained model will be made publicly available on GitHub. %are available at \msh{[GitHub URL]}

\section*{Acknowledgement}

This work was performed for Council for Science, Technology and Innovation (CSTI), Cross-ministerial Strategic Innovation Promotion Program (SIP), “Promoting the application of advanced quantum technology platforms to social issues” (Funding agency : QST).
This work was supported by JSPS KAKENHI Grant-in-Aid for Research Activity Start-up (23K19980).
Classical computational resources were provided by AI Bridging Cloud Infrastructure (ABCI) of National Institute of Advanced Industrial Science and Technology (AIST). 
Quantum computational resources were provided by Amazon Braket.
SM, YS, and TK would like to express the gratitude to Yusuke Hama and Shiro Kawabata for their insightful discussions and supports.
AAG thanks Anders G. Frøseth for his generous support and acknowledges the generous support of Natural Resources Canada and the Canada 150 Research Chairs program.

\section*{Author contributions}
AAG, KN, and TK conceived the conceptual ideas of this study. SM, YS, and TK deviced and outlined the project.
SM implemented the machine-learning algorithms and conducted the experiments with the support of KN, YS, and TK. 
SM and KN wrote the manuscript, and all authors have edited, read and approved the final manuscript.

\bibliographystyle{ieeetr}
\bibliography{main}

%%% For arXiv ------------------------------------------
\newpage
{\centering
    \LARGE \bf \noindent Supplementary Information \\[1em]
}
\vspace{0.5em}
\suppressfloats
\appendix
\renewcommand{\thefigure}{S\arabic{figure}}
\renewcommand{\thetable}{S\arabic{table}}
\renewcommand{\theequation}{S\arabic{equation}}
\setcounter{figure}{0}
\setcounter{table}{0}
\setcounter{equation}{0}

\section*{GQCO model architecture and hyperparameter settings}
\begin{figure}[p]
    \centering
    \includegraphics[clip, width=0.95\columnwidth]{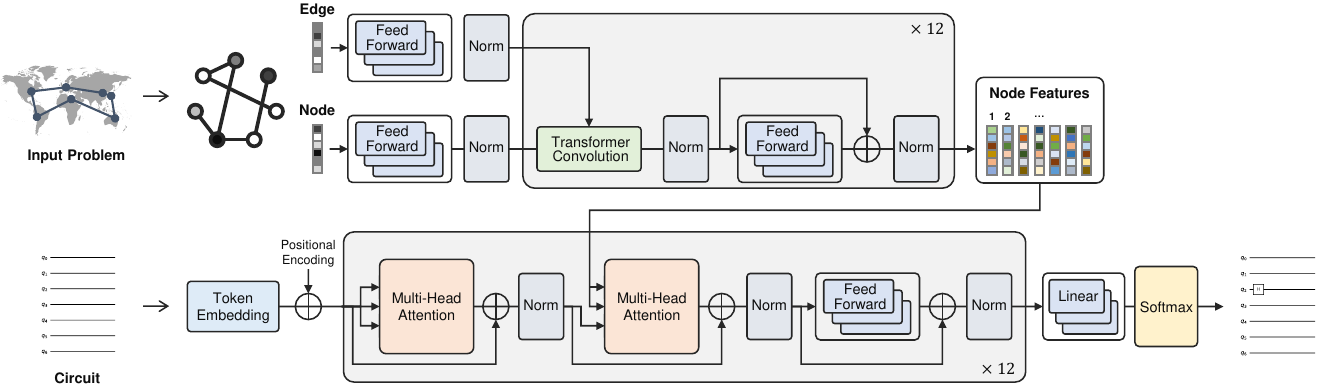}
    \caption{
        Model archtecture.
        As shown in the Results section in the main text, the input combinatorial optimization problem is converted into a graph representation, and the features are engineered. 
        The module containing the Graph Transformer layer is iterated 12 times, and the resulting node features, arranged in node order, serve as the encoding representation. 
        The decoder resembles the structure used in GPT models: \texttt{FeedForward} refers to two linear layers with an activation function between them, and \texttt{Norm} refers to the layer normalization layer. 
        The four \texttt{FeedForward} modules and one linear layer module are organized in a mixture-of-experts (MoE) structure.
    }
    \label{fig:SI-model}
\end{figure}

\suppressfloats
Fig.~\ref{fig:SI-model} illustrates the model structure, and Table~\ref{tab:SI-param} lists the hyperparameter settings used in this work. 
As mentioned in the main text, the entire model consists of 256 million parameters, with 127 million allocated to the encoder and 129 million to the decoder. 
Note that this total number includes parameters for all expert modules up to size 20. 
The number of parameters utilized for a specific problem size is approximately 24 million, with 11 million in the encoder and 13 million in the decoder.

During DPO training, parameters are updated to increase the probability of generating the most preferred circuit. 
Sampling a large number of circuits at once increases the probability of obtaining circuits near the ground state, thereby enhancing training efficiency.
In this study, for each number of qubits $n$, the value of $M$ was determined to maximize GPU memory usage.
These settings are also detailed in Table~\ref{tab:SI-param}.

\begin{table}[p]
    \centering
    \caption{
        Hyperparameter settings for model architecture and training.
        $n$ represents the problem size, i.e., the number of qubits used in the circuit to be generated.
        For the hyperparameters that are not listed, the default settings of PyTorch, PyTorch Geometric, and PyTorch Lightning were used.
    }
    % \small
    \label{tab:SI-param}
    
    \begin{tabular}{lll}
        \hline
             & Hyperparameter & Value \\ 
        \hline
            Base settings &
              Activation function                & GELU               \\
            & Dropout rate                       & 0.0                \\
            & $\epsilon$ for layer normalization & $1 \times 10^{-5}$ \\
        \hline
            Encoder &
              Depth                                            & 12   \\
            & Number of attention heads                        & 8    \\
            & Dimension of the intermediate representations    & 256  \\
            & Dimension of the intermediate feed-forward layer & 1024 \\
        \hline
            Decoder &
              Depth                                            & 12   \\
            & Number of attention heads                        & 8    \\
            & Dimension of the intermediate representations    & 256  \\
            & Dimension of the intermediate feed-forward layer & 1024 \\
        \hline
              Training settings
            & Optimizer            & Adam \\
            & Learning rate        & $1 \times 10^{-4}$ \\ 
            & $(\beta_1, \beta_2)$ for Adam optimizer & $(0.9, 0.95)$ \\
            & $\beta$ for DPO      & 0.1                \\
            & Evaluation frequency & 500                \\
        \hline
              Generation settings
            & Maximum sequence length & $2n$ \\
            & Vocabulary size & $4n^2+15+1$ \\ 
            & Number of samplings during training $M$ for $n=3$ & 1024 \\
            & \phantom{Number of samplings during training $M$} for $n=4$ & 1024 \\
            & \phantom{Number of samplings during training $M$} for $n=5$ & 512 \\
            & \phantom{Number of samplings during training $M$} for $n=6$ & 384 \\
            & \phantom{Number of samplings during training $M$} for $n=7$ & 256 \\
            & \phantom{Number of samplings during training $M$} for $n=8$ & 192 \\
            & \phantom{Number of samplings during training $M$} for $n=9$ & 128 \\
            & \phantom{Number of samplings during training $M$} for $n=10$ & 96 \\
        \hline
    \end{tabular}
\end{table}

\section*{Training history}

\begin{figure}[t]
    \centering
    \includegraphics[clip, width=0.95\columnwidth]{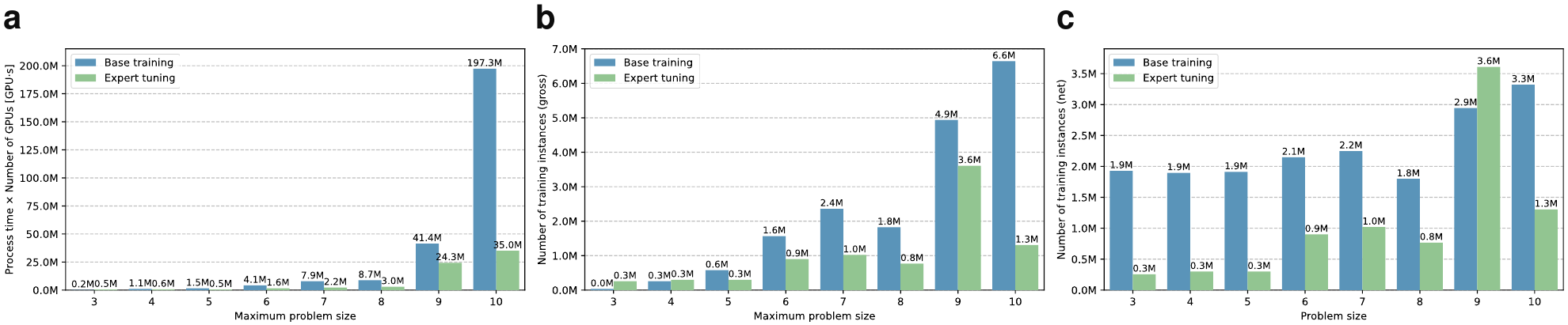}
    \caption{
        Computational costs for training across problem sizes.
        Blue bars represent the costs for base training phase where all the weights of the model are optimized, while green bars indicate the costs for expert tuning phase where each expert in the MoE layer is fine-tuned after the base training.
        (a) 
        Total computational time for each curriculum learning stage.
        The value is calculated by multiplying the process time for distributed training by the number of GPUs utilized during the corresponding stage.
        (b) 
        Gross amount of training data across problem size.
        The total number of training data points is determined by multiplying the number of GPUs by the number of training epochs. 
        The value at each maximum problem size represents the total data generated for that stage, including all problem sizes up to the corresponding size.
        (c) 
        Net amount of training data across problem size.
        This value is derived by adjusting the values of (b) to reflect the net number of data points for each problem size, based on the selection probability Eq.~\eqref{eq:size-probability} of each size.
    }
    \label{fig:SI-training}
\end{figure}

A multi-GPU strategy was used to train the model in a distributed environment. 
Hamiltonian coefficients were generated independently on each GPU, losses were computed locally, and gradients were aggregated across all GPUs to update the model parameters. 
The number of GPUs varied based on available computational resources, ranging from 64 to 256 during training. 
The training process took approximately 20 days.

Fig.~\ref{fig:SI-training}a illustrates the total computational cost for each training size stage, calculated by multiplying the training time per GPU by the number of GPUs used. 
This metric represents the process time required to train on a single GPU. 
Fig.~\ref{fig:SI-training}b shows the product of the number of training epochs and the number of GPUs, which corresponds to the total number of Hamiltonian coefficients generated during training, that is, the number of combinatorial optimization problems processed during training. 
Both metrics increase exponentially as the maximum problem size grows. 
As the number of qubits increases, a greater number of training instances are required, and quantum circuit simulations take longer.

It should be noted that, at each problem size stage, problems up to the maximum size are generated based on the probability Eq.~\eqref{eq:size-probability} in the main text. 
As a result, the number of training instances in Fig.~\ref{fig:SI-training}b includes various problem sizes. 
Correcting this, we show the net number of training instances for each specific problem size in Fig.~\ref{fig:SI-training}c. 
This reveals that the number of training examples increases more slowly with problem size, suggesting that curriculum learning and MoE structures may enhance learning efficiency. 
However, the probability of generating each problem size based on Eq.\eqref{eq:size-probability} is not optimized and may produce an excessive number of instances with small problem size.
In other words, fewer instances than those shown in Fig.~\ref{fig:SI-training}c might be sufficient for learning smaller problem sizes, and the observed scaling in Fig.~\ref{fig:SI-training}c may be overly gradual.

\section*{Additional figures}
\subsection*{Enlarged figures for the real device execution}
\begin{figure}[t]
    \centering
    \includegraphics[clip, width=0.98\columnwidth]{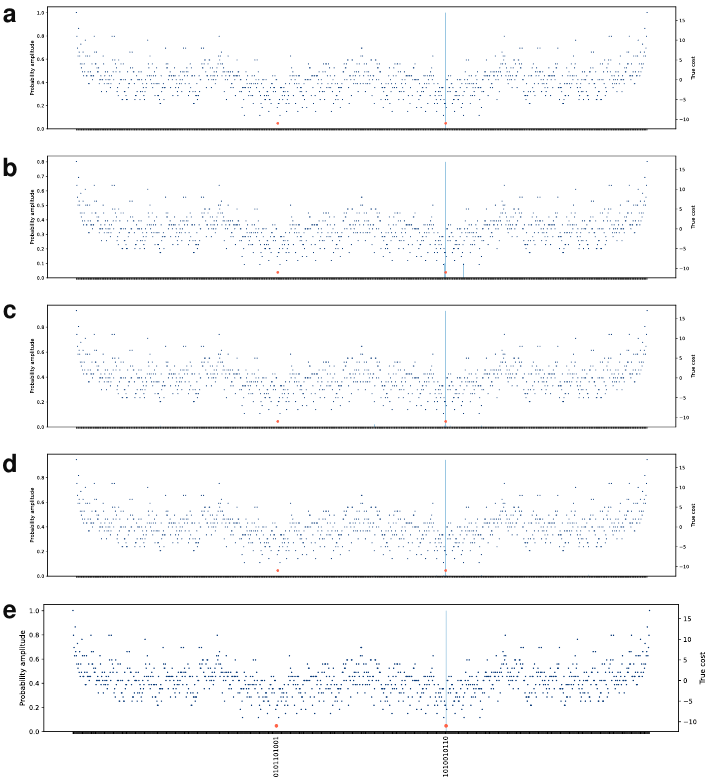}
    \caption{
        Sampling results on the real quantum device for the GQCO-generated circuit.
        (a) 1 shot, (b) 10 shots, (c) 100 shots, (d) 1000 shots, and (e) simulated state vector are displayed.
        The meanings of the plots and histograms are the same as that of Fig.~\ref{fig:mistakes}c.
    }
    \label{fig:SI-GQCO}
\end{figure}

\begin{figure}[t]
    \centering
    \includegraphics[clip, width=0.98\columnwidth]{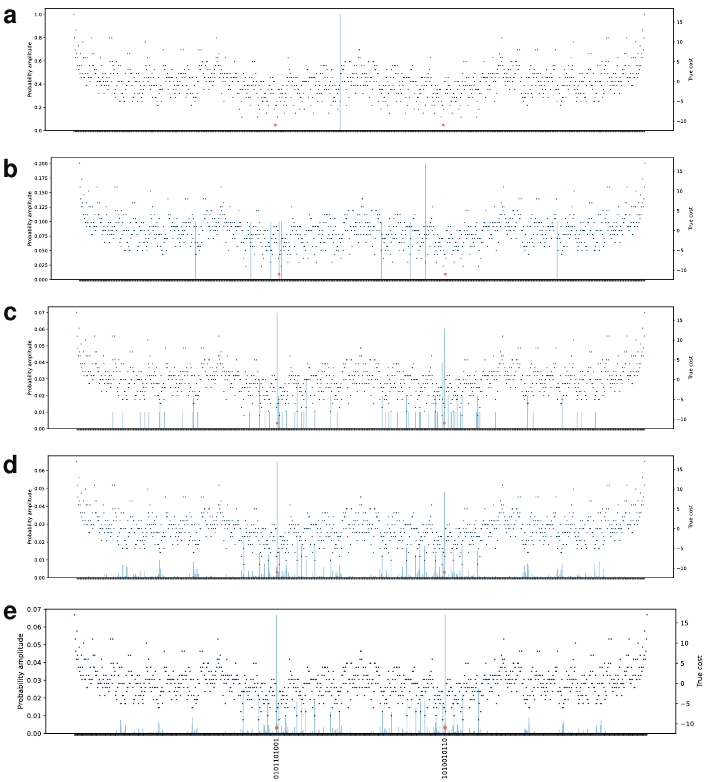}
    \caption{
        Sampling results on the real quantum device for the 2-layer QAOA circuit.
        (a) 1 shot, (b) 10 shots, (c) 100 shots, (d) 1000 shots, and (e) simulated state vector are displayed.
        The meanings of the plots and histograms are the same as that of Fig.~\ref{fig:mistakes}c.
    }
    \label{fig:SI-QAOA}
\end{figure}

The enlarged versions of Fig.~\ref{fig:real}c, displaying the results obtained from the real quantum device, are shown in Fig.~\ref{fig:SI-GQCO} for GQCO and Fig.~\ref{fig:SI-QAOA} for QAOA.

\subsection*{Examples of generated circuits}
\begin{figure}[t]
    \centering
    \includegraphics[clip, width=0.98\columnwidth]{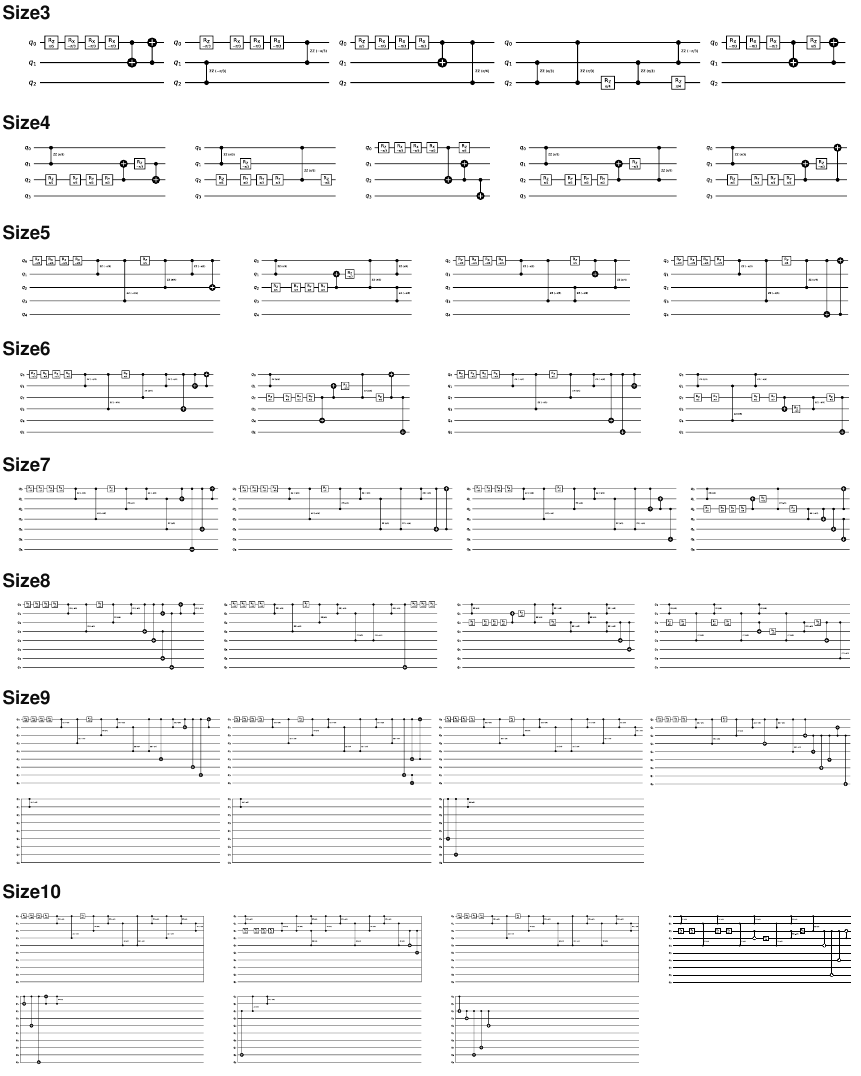}
    \caption{
        Examples of GQCO-generated quantum circuits.
    }
    \label{fig:SI-examples}
\end{figure}

In Fig.~\ref{fig:SI-examples}, we present examples of GQCO-generated circuits for each number of qubits.

%%% ----------------------------------------------------

\end{document}